\documentclass[twocolumn,superscriptaddress,letter]{revtex4}%
\usepackage{amssymb}
\usepackage{color}
\usepackage{graphicx}
\usepackage{dcolumn}
\usepackage{bm}
\usepackage[header,title,page,titletoc]{appendix}
\usepackage{amsmath}
\usepackage{amsfonts}%
\providecommand{\U}[1]{\protect \rule{.1in}{.1in}}

\begin{document}
\title{Topological Hierarchy Insulators and Topological Fractal Insulators}
\author{Jing He}
\affiliation{Department of Physics, Beijing Normal University, Beijing 100875, China}
\affiliation{Department of Physics, Hebei Normal University, HeBei, 050024, P. R. China}
\author{Chun-Li Zang}
\affiliation{Department of Physics, Beijing Normal University, Beijing 100875, China}
\author{Ying Liang}
\affiliation{Department of Physics, Beijing Normal University, Beijing 100875, China}
\author{Su-Peng Kou}
\thanks{Corresponding author}
\email{spkou@bnu.edu.cn}
\affiliation{Department of Physics, Beijing Normal University, Beijing 100875, China}

\begin{abstract}
Topological insulators are new states of quantum matter with metallic
edge/surface states. In this paper, we pointed out that there exists a new
type of particle-hole symmetry-protected topological insulator - topological
hierarchy insulator (THI), a composite topological state of a (parent)
topological insulator and its defect-induced topological mid-gap states. A
particular type of THI is topological fractal insulator, that is a THI with
self-similar topological structure. In the end, we discuss the possible
experimental realizations of THIs.

\end{abstract}
\maketitle

Recently, a new research topic focuses on searching for topologically
nontrivial phases protected by symmetries. Topological band insulator (TBI)
provides an example of symmetry-protected topological states, in which there
exist metallic edge/surface states\cite{Kane,Bernevig}. According to the
characterization of "ten-fold way" of different symmetries\cite{ry}, there are
two types of TBIs in two dimensions: \textrm{Z}-type TBI with integer quantum
Hall effect\cite{thou} and $Z_{2}$-type TBI with quantum spin-Hall
effect\cite{Kane,Roy06,Moore06,Fu06,Essin07}. Then, people found that with
extending the topological classification of band structures to include
considering crystal point group symmetries, there exists new type of TBI -
topological crystalline insulators\cite{fu}. Those additional symmetries lead
to a non-trivial topology of bulk wave functions and gapless edge/surface states.

For TBIs, owning to the "holographic feature", their nontrivial topological
properties can be detected by probing topological defects. For example, in two
dimensional (2D) \textrm{Z}-type TBIs, $\pi$-flux traps mid-gap zero energy
states (zero modes) \cite{Ran,Qi,liu}. For this topological state with a
superlattice of $\pi$-fluxes, due to the polygon rule (Each fermion gains an
accumulated phase shift $\left \vert \phi \right \vert =(\mathrm{n}-2)\pi/2$
encircling around a smallest \textrm{n}-polygon rule \cite{rule}), the
defect-induced mid-gap states can be regarded as an emergent "TBI" with
nontrivial topological properties, including the nonzero Chern number and the
gapless edge states\cite{wu}.

In this paper we point out that a new kind of topological insulator -
\emph{topological hierarchy insulator} may exist, in which the topological
properties are protected by translational and (generalized) particle-hole
symmetry. In this system, the non-topological lattice defect - \emph{vacancy}
traps zero mode\cite{he5}. For the system with vacancy-superlattice, the
ground state becomes topological hierarchy insulator (THI). To show the
nontrivial topological properties of THI, we take the Haldane model on square
lattice and that on honeycomb lattice as examples. We found that the
vacancy-superlattice could induce topological mid-gap states and the low
energy physics of the localized modes become that of an emergent "TBI" on the
parent TBI. In particular, the so-called \emph{topological fractal insulator}
appears as a special THI with \emph{self-similar} topological structure.

\textit{The Haldane model on square lattice}: Our starting point is the
(spinless) Haldane model on square lattice\cite{hal,yuyue}, of which the
Hamiltonian is
\begin{align}
\hat{H}_{\mathrm{parent}}  &  =-t\sum_{i\in A}\left(  \hat{c}_{i+\hat{x}%
}^{\dagger}\hat{c}_{i}+i\hat{c}_{i+\hat{y}}^{\dagger}\hat{c}_{i}+h.c.\right)
\nonumber \\
&  +t\sum_{i\in B}\left(  \hat{c}_{i+\hat{x}}^{\dagger}\hat{c}_{i}-i\hat
{c}_{i+\hat{y}}^{\dagger}\hat{c}_{i}+h.c.\right) \nonumber \\
&  -t^{\prime}\sum_{i\in A}\left(  \hat{c}_{i+\hat{x}+\hat{y}}^{\dagger}%
\hat{c}_{i}+\hat{c}_{i+\hat{x}-\hat{y}}^{\dagger}\hat{c}_{i}+h.c.\right)
\nonumber \\
&  +t^{\prime}\sum_{i\in B}\left(  \hat{c}_{i+\hat{x}+\hat{y}}^{\dagger}%
\hat{c}_{i}+\hat{c}_{i+\hat{x}-\hat{y}}^{\dagger}\hat{c}_{i}+h.c.\right)
\label{1}%
\end{align}
where $\hat{c}_{i}$ is the annihilation operator of the fermions at the site
$i$. $A$ and $B$ label the sublattices. $t$ and $t^{\prime}$ are the nearest
neighbor (NN) and the next nearest neighbor (NNN) hoppings, respectively. For
the Hamiltonian in Eq.(1), there exists $\pi$-flux in each square plaquette
and $\frac{\pi}{2}$-flux in each triangular lattice. In this paper we set the
lattice spacing $a$ to be unit.

The Haldane model on square lattice is a TBI with quantum anomalous Hall (QAH)
effect. There is a finite energy gap near points $\mathbf{k}_{1}=(0,$ $0)$ and
$\mathbf{k}_{2}=(\pi,$ $0)\ $which is $\Delta_{f}=8t^{\prime}$. The Chern
number $C_{\mathrm{parent}}$ of (parent) haldane model is $1$ and the
quantized Hall conductivity is $\frac{e^{2}}{h}$.\

In particular, we point out that the Hamiltonian in Eq.(1) has the
particle-hole (PH) symmetry. Under PH transformation, we have $\hat
{H}_{\mathrm{parent}}=-\mathcal{P}^{\dagger}\hat{H}_{\mathrm{parent}%
}\mathcal{P}$ where $\mathcal{P}=\mathcal{R}\cdot \mathcal{K}$ is the PH
transformation operator\cite{yang,he5}. Here, $\mathcal{R}$ is an operator
that leads to $\hat{c}_{i}\leftrightarrow(-1)^{i}\hat{c}_{i}^{\dagger}$ and
$\mathcal{K}$ is the complex conjugate operator. As a result, each energy
level with positive energy $E$ is paired with an energy level with negative
energy $-E$. Because the Hamiltonian in Eq.(1) is on a bipartite lattice, the
quantum levels of the system with a single vacancy become an odd integer
number. As a result, there must exist an unpaired electronic state when we
remove a lattice-site to create a vacancy. Due to PH symmetry, the
corresponding unpaired electronic state must have exactly zero
energy\cite{he5}. We denote the wave-function of the zero mode by $\psi
_{0}(r_{i}-R)$ where $R$ is the position of the vacancy. The quantum states of
the fermionic zero mode around a vacancy can be formally described in terms of
the fermion Fock states $\left \{  \left \vert 0\right \rangle ,\left \vert
1\right \rangle \right \}  $. Here, $\left \vert 0\right \rangle ,\left \vert
1\right \rangle $ denote the empty state,\ and the occupied state, respectively.

\textit{Topological hierarchy insulator}: We next study the Haldane model with
vacancy-superlattice (VSL). When there are two vacancies nearby, the
inter-vacancy quantum tunneling effect occurs and the fermionic zero modes on
two vacancies couple. The lattice constant of the VSL is denoted by $L$, which
is the distance between two vacancies nearby. When the vacancies locate at the
different sublattices, the energy splitting $\Delta E$ is finite ($L/a$ is an
odd number); when the vacancies locate at the same sublattice, the energy
splitting vanishes, $\Delta E=0$ ($L/a$ is an even number). For the case of
$\xi<L$, we can consider each vacancy as an isolated "atom" with localized
electronic states and use the effective tight-binding model to describe these
quantum states induced by the VSL. Now, we superpose the localized states to
obtain the sets of Wannier wave functions $\psi_{0}(r_{i}-R)$. In general, the
effective tight-binding model of the localized states induced by the VSL
becomes
\begin{align}
\hat{H}_{1-\mathrm{VL}} &  =-\sum_{\left \langle R,R^{\prime}\right \rangle
}T_{RR^{\prime}}\hat{d}_{R}^{\dagger}\hat{d}_{R^{\prime}}-\sum_{\left \langle
\left \langle R,R^{\prime}\right \rangle \right \rangle }T_{RR^{\prime}}^{\prime
}\hat{d}_{R}^{\dagger}\hat{d}_{R^{\prime}}\label{ve}\\
&  -\sum_{\left \langle \left \langle \left \langle R,R^{\prime}\right \rangle
\right \rangle \right \rangle }T_{RR^{\prime}}^{\prime \prime}\hat{d}%
_{R}^{\dagger}\hat{d}_{R^{\prime}}+...,\nonumber
\end{align}
where $\hat{d}_{R}$ is the fermionic annihilation operator of a localized
state on a vacancy $R$. $T_{RR^{\prime}}$ ($T_{RR^{\prime}}^{\prime},$
$T_{RR^{\prime}}^{\prime \prime}$) is the hopping parameter between NN (NNN,
NNNN) sites $R$ and $R^{\prime}$ and is determined by the energy splitting
$\Delta E_{RR^{\prime}}$, $\left \vert T_{RR^{\prime}}\right \vert =\left \vert
\Delta E\right \vert $ ($\left \vert T_{RR^{\prime}}^{\prime}\right \vert
=\left \vert \Delta E^{\prime}\right \vert ,$ $\left \vert T_{RR^{\prime}%
}^{\prime \prime}\right \vert =\left \vert \Delta E^{\prime \prime}\right \vert $).
In general, due to $\left \vert T_{RR^{\prime}}^{\prime \prime}\right \vert
\ll \left \vert T_{RR^{\prime}}^{\prime}\right \vert ,$ $\left \vert
T_{RR^{\prime}}\right \vert $, we only consider the NN hopping term and NNN
hopping term. Because the parent Hamiltonian in Eq.(1) has PH symmetry, the
effective tight-binding model of the localized states induced by the VSL
$\hat{H}_{1-\mathrm{VL}}$ in Eq.(\ref{ve}) also has PH symmetry, $\hat
{H}_{1-\mathrm{VL}}=-\mathcal{P}^{\dagger}\hat{H}_{1-\mathrm{VL}}\mathcal{P}$.

Because the energy splitting $\Delta E_{RR^{\prime}}$ only determines the
particle's hopping amplitude $\left \vert T_{RR^{\prime}}\right \vert $, we need
to settle down the phase of $T_{RR^{\prime}}$. In particular, to preserve PH
symmetry, the total phase around a plaquette with four vacancies can be either
$\pi$ or $0$. An approach is to count the total flux number inside the
plaquette. For the mid-gap states induced by VSL, every four vacancies form an
$L\times L$ square, of which the total flux number is just $L^{2}/2$ and the
total phase around a plaquette is $\pi L^{2}$. After considering the compact
condition, we have $0/\pi$-flux inside each plaquette if $L^{2}$ is an
even/odd number. In addition, we also use a numerical approach to check above
prediction of the flux number inside the plaquette.

\begin{figure}[ptbh]
\includegraphics[width = 8 cm]{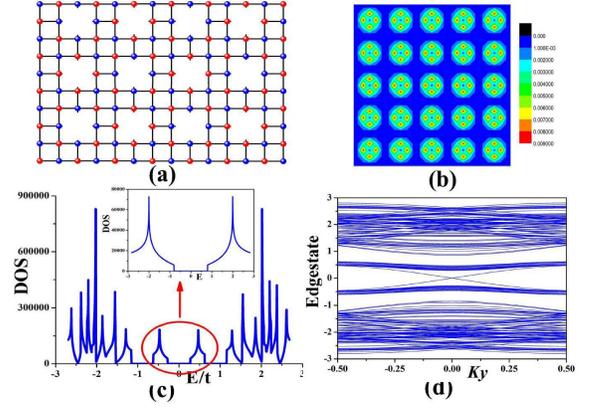}\caption{(Color online) (a) The
illustration of square-VSLs on square lattice; (b) The particle distribution
of zero modes around vacancies; (c) The DOS of the Haldane model with
square-VSL for $L=3a$: the mid-gap states are induced by the square-VSL. The
inset shows the DOS of the parent Haldane model; (d) The edge states of
mid-gap states. The parameter is $t^{\prime}/t=0.2.$}%
\end{figure}

Based on the Haldane model, we study the properties of a PH symmetry-protected
TBI with VSL and introduce the concept of \emph{topological hierarchy
insulator}. To make it clear, we take the Haldane model with \emph{square}-VSL
as an example, of which the lattice constant $L$ is $3a$. See the illustration
in Fig.1.(a). From above discussion, there exists $\pi$-flux in each square
plaquette of VSL and $\frac{\pi}{2}$-flux in each triangular VSL. Fig.1.(b) is
the configuration of zero modes around vacancies. From Fig.1.(b), one can see
that quantum states induced by VSL can be regarded as a new generation of
lattice model, of which the vacancies play the role of the "atoms".

As a result, we obtain an effective Hamiltonian to describe the
vacancy-induced mid-gap states, that reads
\begin{align}
\hat{H}_{1-\mathrm{VL}}  &  =-T\sum_{R\in \bar{A}}(\hat{d}_{R+L\hat{x}%
}^{\dagger}\hat{d}_{R}+i\hat{d}_{R+L\hat{y}}^{\dagger}\hat{d}_{R}%
+h.c.)\label{vl}\\
&  +T\sum_{R\in \bar{B}}(\hat{d}_{R+L\hat{x}}^{\dagger}\hat{d}_{R}-i\hat
{d}_{R+L\hat{y}}^{\dagger}\hat{d}_{R}+h.c.)\nonumber \\
&  +T^{\prime}\sum_{i\in \bar{A}}(\hat{d}_{R+L\hat{x}+L\hat{y}}^{\dagger}%
\hat{d}_{R}+\hat{d}_{R+L\hat{x}-L\hat{y}}^{\dagger}\hat{d}_{R}%
+h.c.)\nonumber \\
&  -T^{\prime}\sum_{i\in \bar{B}}(\hat{d}_{R+L\hat{x}+L\hat{y}}^{\dagger}%
\hat{d}_{R}+\hat{d}_{R+L\hat{x}-L\hat{y}}^{\dagger}\hat{d}_{R}+h.c.)\nonumber
\end{align}
where $T,$ $T^{\prime}$ are effective NN and NNN hopping parameters,
respectively. $\bar{A}$ and $\bar{B}$ label the sublattices of VSL. By
comparing Eq.(\ref{1}) and Eq.(\ref{vl}), one can see the corresponding
relationship, $\hat{c}_{i}\longleftrightarrow \hat{d}_{R}$,
$t\longleftrightarrow T,$ $t^{\prime}\longleftrightarrow-T^{\prime}$. For the
defect-induced quantum states, the energy spectrum and DOS are obviously
similar to those of the parent Hamiltonian in Eq.(1).

In Fig.1.(c) we show the DOS of the Haldane model with square-VSL for
$t^{\prime}/t=0.2$ and $L=3a$. In Fig.1.(c), the mid-gap (MG) energy bands
appear and the DOS of the MG states is really similar to that of the parent
TBI (the inset in Fig.1.(c)). We obtain the hopping parameters by fitting the
DOS of the defect-induced quantum states as $T=0.22t$ and $T^{\prime}=0.07t.$

In addition, we will show the MG states induced by VSL have similar
topological properties to those of parent TBI. By the numerical approach, we
found that the Chern number of the MG states is $C_{1-\mathrm{VL}}=-1$ which
opposites to that of parent TBI as $C_{1-\mathrm{VL}}=-C_{\mathrm{parent}}$.
To characterize the THI, we define the filling factor, $\nu=N_{f}/N$ where $N$
is the total number of lattice sites and $N_{f}$ is the number of fermions.
The composite system is an insulator at filling factor $\nu=\frac{1}{2}$ or
$\nu=\frac{1}{2}(1-\frac{a^{2}}{L^{2}})$. At half filling (or $\nu=\frac{1}%
{2}$), the total Chern number of the system is zero due to $C_{\mathrm{total}%
}=C_{\mathrm{parent}}+C_{1-\mathrm{VL}}$; at filling factor $\frac{1}%
{2}(1-\frac{a^{2}}{L^{2}})$, the total Chern number of the system is $1$ due
to $C_{\mathrm{total}}=C_{\mathrm{parent}}$.

To illustrate the topological properties of the Haldane model with square-VSL,
we then study its edge states. On one hand, at the filling factor $\frac{1}%
{2}(1-\frac{a^{2}}{L^{2}})$, the system becomes a TBI with Chern number $1$.
As a result, when the system has periodic boundary condition along
$y$-direction but open boundary condition along $x$-direction, there exist the
gapless edge states. On the other hand, we consider the half filling case, of
which the system has zero Chern number. However, the system still has
nontrivial topological properties. When the parent topological insulator has
periodic boundary condition along both $x$-direction and $y$-direction, while
the VSL has periodic boundary condition along $y$-direction but open boundary
condition along $x$-direction, there may exist gapless edge states. See the
numerical results in Fig.1.(d) for the case of $L=3a$. We can see that there
indeed exist gapless edge states on the boundaries of VSL.

In general, the Haldane model with a square-VSL of odd number $L^{2}$ has
nontrivial topological properties and is different from traditional TBIs. As a
result, we call this type of PH symmetry-protected TBI with topological MG
states to be\emph{ topological hierarchy insulator}. In other words, the point
of view THI is given by the following equation
\begin{align}
\text{THI }  &  =\text{ PH symmetry-protected TBI }\nonumber \\
&  \text{+ }L\times L\text{-square-VSL}\nonumber \\
&  =\text{ Parent TBI + topological MG states.}%
\end{align}

\textit{Topological fractal insulator}: We then study a particular type of THI
- topological fractal insulator (TFI), that is a THI with infinite generations
of self-similar $L\times L$-square-VSLs. See the illustration in Fig.2.(a).

\begin{figure}[ptbh]
\includegraphics[width =9cm]{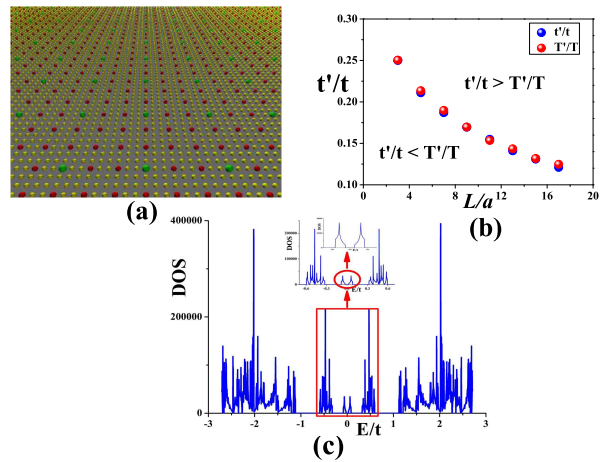}\caption{(Color online) (a) An
illustration of the VSLs for TFI; (b) The self-similar condition; (c) The DOS
of TFI. The parameter is $t^{\prime}/t=0.2$. The insets show the DOSs of the
mid-gap states.}%
\end{figure}

To construct a TFI, firstly we consider a THI with self-similar MG states.
Because we can tune $T^{\prime}/T$ (or $\Delta E^{\prime}/\Delta E$) by
changing the NNN hopping $t^{\prime}$ of the parent TBI or the VSL constant
$L$, the MG states can be self-similar to the parents states for the case of
$T/T^{\prime}=t/t^{\prime}$. After numerical calculations, we obtain the
self-similar condition in Fig.2.(b). See the dots in Fig.2.(b). According to
the phase diagram of Fig.2.(b), smaller/bigger $t/t^{\prime}$ leads to
bigger/smaller $T/T^{\prime}$. Now, under the self-similar condition
$T/T^{\prime}=t/t^{\prime}$, the effective Hamiltonian that describes the MG
states becomes $\hat{H}_{1-\mathrm{VL}}=\alpha$ $\hat{H}_{\mathrm{parent}%
}^{\ast}$ where $\alpha=T/t$ ($\alpha<1$) is an energy-scaling ratio.

Next, we regard the vacancy-induced MG states as a parent TBI and consider an
additional VSL on it. So, using the recursive relation, we can see that there
exist additional MG states inside the energy gap of first generation MG states
(we call it generation-$1$ MG states). Consequently, we get a new generation
of topological MG states (we call such MG states in the energy gap of MG
states as generation-$2$ MG states). Using the same method, we can construct a
THI with $n$-generation MG states and eventually a THI with infinite
generations of self-similar MG states that is just a TFI. Then, under the
self-similar condition $T/T^{\prime}=t/t^{\prime}$, the effective Hamiltonian
of the generation-$n$ MG states is given by
\begin{equation}
\hat{H}_{n-\mathrm{VL}}=\{%
\begin{array}
[c]{c}%
\alpha^{n}\text{ }\hat{H}_{\mathrm{parent}}\text{, }n\text{ is an even
number,}\\
\alpha^{n}\text{ }\hat{H}_{\mathrm{parent}}^{\ast}\text{, }n\text{ is an odd
number.}%
\end{array}
\end{equation}
From the DOS in Fig.2.(c), we can see that the DOS of the MG states is always self-similar.

In addition, we discuss the Chern number of the TFI. The TFI is an insulator
at filling factor $\nu=\frac{1}{2}(1-\frac{a^{2m}}{L^{2m}})$ where $m$ is an
integer number, $m\geq1$. The Chern number of generation-$m$ MG states is
known to be $C_{m-\mathrm{VL}}=(-1)^{m}$. As a result, when the filling factor
is $\nu=\frac{1}{2}(1-\frac{a^{2m}}{L^{2m}})$, the total Chern number of the
system is
\begin{equation}
C_{\mathrm{total}}=C_{\mathrm{parent}}+\sum_{n=1}^{m-1}C_{n-\mathrm{VL}%
}=[1+(-1)^{m-1}]/2.
\end{equation}
For example, when the filling factor is $\frac{1}{2}(1-\frac{a^{6}}{L^{6}})$
($m=3$), the total Chern number of the system is $1$ due to $C_{\mathrm{total}%
}=C_{\mathrm{parent}}+C_{1-\mathrm{VL}}+C_{2-\mathrm{VL}}.$

It is known that a self-similar object looks "roughly" the same on different
scales and fractal is a particularly self-similar object that exhibits a
repeating pattern displaying at every scale. Topological fractal insulator
provides a unique example of topological matters with self-similarity, in
which the topological properties always look the same on different energy
scales ($\alpha^{2n}$) or different length scales ($L^{2n}$).

\textit{Topological hierarchy insulators on honeycomb lattice}: We finally
study the THI based on the Haldane model on honeycomb lattice, of which the
Hamiltonian is given by
\begin{equation}
\hat{H}_{\mathrm{parent}}=-t\sum_{\left \langle i,j\right \rangle }\hat{c}%
_{i}^{\dagger}\hat{c}_{j}-t^{\prime}\sum \limits_{\left \langle \left \langle
{i,j}\right \rangle \right \rangle }e^{i\phi_{ij}}\hat{c}_{i}^{\dagger}\hat
{c}_{j}-\mu \sum_{i}\hat{c}_{i}^{\dagger}\hat{c}_{i}\label{haldane}%
\end{equation}
where $\hat{c}_{i}$ represents the fermion annihilation operator at site $i.$
$t$ and $t^{\prime}$ are the nearest neighbor (NN) and the next nearest
neighbor (NNN) hoppings, respectively. $e^{i\phi_{ij}}$ is a complex phase
along the NNN link, and we set the direction of the positive phase is
clockwise $\left(  \left \vert \phi_{ij}\right \vert =\frac{\pi}{2}\right)  $.
$\mu$ is the chemical potential. The Haldane model on honeycomb lattice is
also a Chern insulator with nonzero Chern number, $C_{\mathrm{parent}}=1$.\

Since the Haldane model of Eq.(\ref{haldane}) also satisfies the PH symmetry
\cite{he5}, $\hat{H}_{\mathrm{parent}}=-\mathcal{P}^{\dagger}\hat
{H}_{\mathrm{parent}}\mathcal{P}$, a vacancy induces a fermionic zero mode. In
the followings, we focus on the Haldane model with honeycomb-VSL shown in
Fig.3.(a), of which the distance between two vacancies with shortest distance
$L$ is $5a.$ The particle density distribution of fermionic zero modes induced
by VSL is shown in Fig.3.(b).

\begin{figure}[ptbh]
\includegraphics[width =8cm]{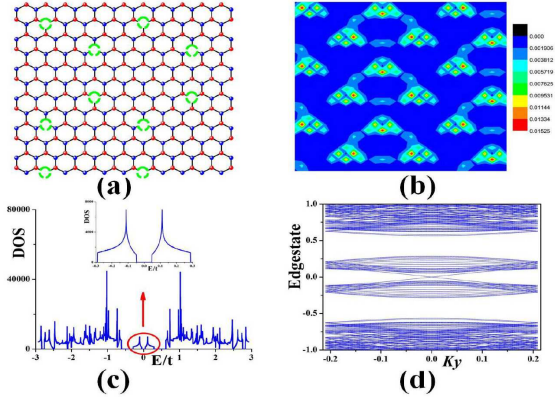}\caption{(Color online) (a) The
illustration of honeycomb-VSL on honeycomb lattice. The distance between two
vacancies is $5a$; (b) The particle distribution of zero modes around
vacancies; (c) The DOS of the Haldane model with honeycomb-VSL. The insets
show the DOSs of the mid-gap states; (d) The edge states of mid-gap states.
The parameter is $t^{\prime}/t=0.1$.}%
\end{figure}

Using similar approach, we derive an effective Hamiltonian to describe the MG
states
\begin{equation}
\hat{H}_{\mathrm{I-VL}}=-T\sum_{\left \langle I,J\right \rangle }\hat{d}%
_{I}^{\dagger}\hat{d}_{J}-T^{\prime}\sum \limits_{\left \langle \left \langle
I{,J}\right \rangle \right \rangle }e^{-i\phi_{IJ}}\hat{d}_{I}^{\dagger}\hat
{d}_{J}-\mu \sum_{I}\hat{d}_{I}^{\dagger}\hat{d}_{I} \label{h}%
\end{equation}
where $T,$ $T^{\prime}$ are effective NN and NNN hopping parameters,
respectively. $I$, $J$ denote the positions of a vacancies. By comparing
Eq.(\ref{haldane}) and Eq.(\ref{h}), one can see the corresponding
relationship to be $\hat{c}_{i}\longleftrightarrow \hat{d}_{I}$,
$t\longleftrightarrow T,$ $t^{\prime}\longleftrightarrow T^{\prime}$,
$\phi_{ij}\rightarrow-\phi_{IJ}.$

The DOS of the THI on honeycomb lattice is shown in Fig.3.(c). There exist MG
states induced by honeycomb-VSL, of which the DOS is also similar to that of
the parent Hamiltonian in Eq.(\ref{haldane}). Next, we calculate the Chern
number of the THI on honeycomb lattice and find that the Chern number of the
MG states $C_{\mathrm{I-VL}}$ is $-1$ which also opposites to the parent TBI,
$C_{\mathrm{parent}}=-C_{\mathrm{I-VL}}$. The THI is an insulator at filling
factor $\nu=\frac{1}{2}$ or $\nu=\frac{24}{50}$. At half filling (or
$\nu=\frac{1}{2}$), the total Chern number of the system is zero; at filling
factor $\nu=\frac{24}{50}$, the total Chern number of the system is $1$. In
addition, we study the edge states of the MG states. See the numerical results
in Fig.3.(d). For a THI with $n$-generation MG states, the total Chern number
of the system at filling factor $\nu=\frac{1}{2}(1-\frac{a^{2n}}{L^{2n}})$ is
$[1+(-1)^{n-1}]/2.$

\textit{Discussion and conclusion}: In this paper, based on the Haldane model
on square lattice and that on honeycomb lattice, we studied the Chern
insulators with VSL and found the vacancy-induced MG states have nontrivial
topological properties. Then, we discover new types of TBI - THIs and TFIs.
All these topological insulators are protected by particle-hole symmetry. In
particular, TFIs provide a unique example of topological matters with
self-similarity, in which the topological properties always look the same on
different energy scales. As a result, these exotic properties of THIs and TFIs
will deepen our understanding of symmetry-protected topological states. In the
future, similar topological states (THIs or TFIs) can be explored on other
lattices (for example, the triangle lattice or kagome lattice).

Finally, we address the relevant experimental realizations of the THI. In
condensed matter physics, it is difficult to realize the Haldane model.
However, the Kane-Mele (KM) model can be regarded as two copies of the Haldane
model such that the spin up electron and the spin down electron exhibit
opposite Chern nmber\cite{Kane}. Because the KM model\ also has PH symmetry,
THI can be designed based on it. Ref.~\cite{yao} proposed that the KM model
can be realized in a silicene-based material with a honeycomb lattice. A
possible vacancy in silicene is a missing Si-atom that can be regarded as a
vacancy in the KM model. For the silicene with periodic pattern of missing
Si-atoms, the THI\ may be possible realized. On the other hand, experimental
realizations of quantum many-body systems in optical lattices have led to a
chance to simulate a variety of topological states. Recently, in
Ref.\cite{jot}, the (spinless) Haldane model on honeycomb optical lattice has
been simulated in the cold atoms. Because the VSL on optical lattice may be
achieved by holographic method, THI may be realized on optical lattice by
using timely cold-atom technology.\

\begin{center}
{\textbf{* * *}}
\end{center}

This work is supported by National Basic Research Program of China (973
Program) under the grant No. 2011CB921803, 2012CB921704 and NSFC Grant No.
11174035, 11474025, 11404090, Natural Science Foundation of Hebei Province, Hebei Education Department Natural Science Foundation, SRFDP.

\end{document}